\documentclass{ws-procs9x6}
\usepackage{epsfig,subfigure}
\usepackage{cite}
\begin{document}

\title{Observations of X-ray flares and associated MHD oscillations in star $\xi$ Boo}

\author{J. C. Pandey$^*$ and A. K. Srivastava}

\address{Aryabhatta Reearch Institute of Observational Sciences,\\
Manora Peak, Nainital 263139, India\\
$^*$E-mail: jeewan@aries.res.in\\
www.aries.res.in}

\begin{abstract}
Using data from observations made with XMM–Newton, we present an X-ray analysis of two flares observed in $\xi$ Boo. The  flare loop parameters are derived using various loop models including state-of-the-art hydrodynamic flare model. The loop lengths derived for the flaring loop structure  are found to be  less than the stellar radius. The exponential decay of the X-ray light curves, and time evolution of the plasma temperature and emission measure are similar to those observed in compact solar flares.  The X-ray light curve of post flare phase is investigated with wavelet analysis. Wavelet  analysis clearly show oscillations of the period of 1019 s.  Using the observationally estimated loop length, density and magnetic field, the theoretically derived oscillation  period for fast-kink mode approximately matches with the observationally  estimated period.  This is the first likely  observational evidence of fundamental fast-kink mode of magnetoacoustic waves in the stellar loops during the post-flare phase of heightened emission.

\end{abstract}


\bodymatter

\vspace{-1mm}
\section{Introduction}

Sudden enhance and gradual decay in the intensity of a star is the most known feature of a flare observed in it.  During the flare large amount of energy is released in a short interval of time. Flares are known to be a manifestation of the reconnection of magnetic loops, accompanied by particle beams, chromospheric evaporation, rapid bulk flows or mass ejection, and heating of plasma confined in loops. Stellar flares could be radiating several orders of magnitude more energy than a solar flare. The flare produced by the stellar sources (e.g. RS CVn and BY Dra types) are usually detected in  UV and X-rays.  Flares from these stars present many analogies with the solar flares. However, they also show significant differences, such as the amount of energy released.  Modeling the dynamic behavior of X-ray flares allows us to constrain the properties of magnetic loops in ways that cannot be done from analysis of quiescent coronae that provide only a spatial and temporal average over some ensemble of structures. Therefore, study of stellar flares is a valuable tool for understanding stellar coronae, as these are dynamical events, which therefore embody different information than available from static, time-averaged observations (e.g. \cite{pandey09}, \cite{pandey08}, \cite{favata05}, \cite{reale04}) .

With the advent of HINODE/STERIO the study of oscillations in the solar corona became  the most interesting now a days. The waves in the solar corona have been explained by magneto-hydrodynamic (MHD) oscillations in coronal loops \cite{roberts00}. MHD waves and oscillations in the solar and  probably in the stellar atmospheres are assumed to be generated by coupling of complex magnetic field and plasma.  These MHD waves and oscillations are one of the important candidates for heating  the solar/stellar coronae and accelerating the supersonic winds. In the Sun, magnetically structured flaring  loops, anchored into the photosphere, exhibit various kinds of MHD
oscillations (e.g. fast sausage, kink and slow acoustic oscillations). Stellar MHD seismology is also a new developing tool to deduce the physical properties of the atmosphere of magnetically active stars, and it is based on the analogy
of the solar coronal seismology \cite{nakariakov07}.
One of the most exciting aspects of "coronal seismology" is that it potentially provides us with the capability for determining the magnetic field strength in the corona \cite{roberts84}, as well as, in the stellar case, with information on otherwise unresolved spatial scales, e.g., flare loop lengths.

In this article, we have studied the X-ray flares and associated MHD oscillations in the post flare phase of the star $\xi$ Boo.  $\xi$ Boo is a nearby (6.0 pc) visual binary, comprising a primary G8 dwarf and a secondary K4 dwarf with an orbital period of 151 yr. In terms of the outer atmospheric emission, UV and X-ray observations show that the primary dominates entirely over the secondary \cite{laming99}. Recently, using the Chandra observations Wood \& Linsky \cite{wood10} show that $\xi$ Boo primary contributes 88.5\% of total X-ray emission.

\section{Observations and data reduction}
$\xi$ Boo was observed with XMM-Newton satellite using the EPIC MOS detectors and Reflection Grating Spectrometer (RGS) on 2001 January 19 at 19:25:06 UT for 59 ks. The details on the XMM-Newton satellite and its detectors are given in the references \cite{jansen01} and \cite{struder01}.  The data were reduced  using the Science Analysis System  version 8.0 with updated calibration files. Details of data  reduction is given in \cite{pandey08}. The stellar separation of xi Boo primary and xi Boo secondary is 6.3 arcsec, which more than the spatial  resolution of XMM-Newton. Therefore, we can safely consider that $\xi$ Boo primary is observed by XMM-Newton.

\section{X-ray light curves}

\begin{figure}
\begin{center}
\includegraphics[width=2.0in,angle=-90]{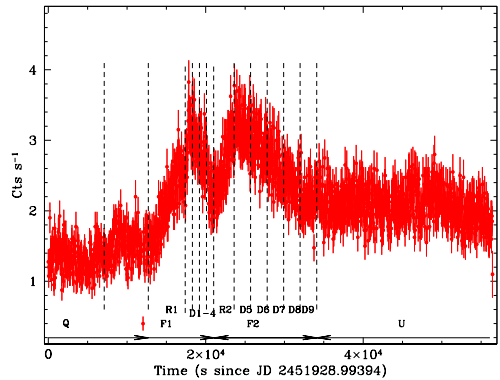}
\caption{X-ray light curve of the star $\xi$ Boo detected by MOS1, vertical lines are different time segments.}
\label{fig:mos_lcurve}
\end{center}
\end{figure}

The MOS light curve of $\xi$ Boo in the energy band 0.3-10 keV is shown in Figure \ref{fig:mos_lcurve}. Two flare like features, marked as F1 and F2 are clearly seen. The quiescent state in the light curve is marked by Q. A higher emission marked by U is also seen in the X-ray light curve. Both flares are found to be $\sim 2.6$ times more energetic than that of the quiescent state. A flux of $2\times 10^{-11}$ erg cm$^{-2}$s$^{-1}$ was observed during the quiescent state of $\xi$ Boo. The e-folding decay times of the flares F1 an F2 were found to be $4396\pm333$ and $9885\pm362$ s, respectively. However, the e-folding rise times were found to be $3169\pm259$ and $3351\pm244$ s for the flares F1 and F2, respectively. The duration of flares F1 and F2 were 8.4 and 11.3 ks. From the light curve, it is seen that  before ending the  flare F1 the flare F2 starts. Similar loop systems have been observed in a flare on the Sun (e.g. so-called Bastille Day flare \cite{aschwanden01}), and in a stellar analogue dMe star Proxima Centauri \cite{reale04}. In these two events, a double ignition in nearby loops was observed or suggested,
and the delay between the ignitions appears to scale with the loop sizes.
Both flares are appeared to be a long decay flare ($\tau_d \geq1$ h).

\section{Evolutions of flares}
\begin{figure}
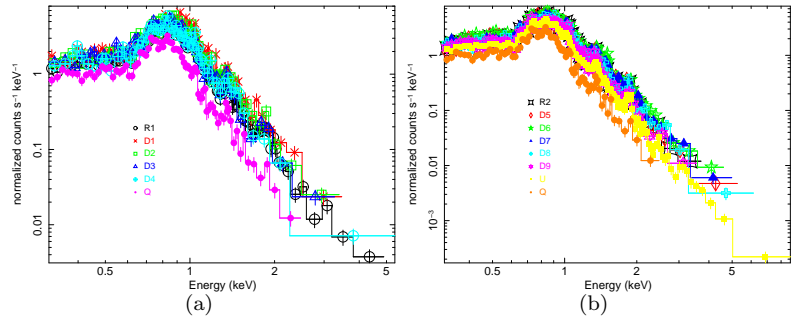

\begin{center}
\subfigure[]{\includegraphics[width=1.5in,angle=-90]{xiboo_spec_f1.ps}}
\subfigure[]{\includegraphics[width=1.5in,angle=-90]{xiboo_spec_f2.ps}}
\caption{(a) Spectra of different time segments during the flare F1, and (b) Spectra of different time segments  during the flare F2. The quiescent state spectra is also shown by solid circles in  each Figure (a) and (b)}
\label{fig:spectra}
\end{center}
\end{figure}

In order to trace the spectral changes during the flares, we have
analysed the spectra of the different time intervals shown in Fig. \ref{fig:mos_lcurve}.  Flare F1 and F2 have binned into five and six intervals, respectively. The  spectra  corresponding to different time segments associated with flares F1 and F2 are shown in Fig. \ref{fig:spectra} (a)  and \ref{fig:spectra} (b), respectively. The quiescent state spectra are also shown by solid circles in each Fig. \ref{fig:spectra} (a) and \ref{fig:spectra}(b) for comparison. The quiescent state spectra of $\xi$ Boo were fitted with a single (1T) and two (2T) temperature collisional plasma model known as APEC \cite{smith01}, with variable elemental abundances. Only 2T plasma model with variable abundance ($Z < 0.23 Z_\odot$) were found acceptable. The temperature and emission measure (EM) were found to be $0.20\pm0.05$ keV and $15\pm6\times 10^{51}$ cm$^{-3}$ for cool component, and $0.57_{}^{}$ keV and $12\pm 1 \times 10^{51}$ cm$^{-3}$, respectively. The luminosity during the quiescent state was found to be $5\times10^{28}$ erg s$^{-1}$.

\begin{figure}
\begin{center}
\subfigure[]{\includegraphics[width=1.0in,angle=-90]{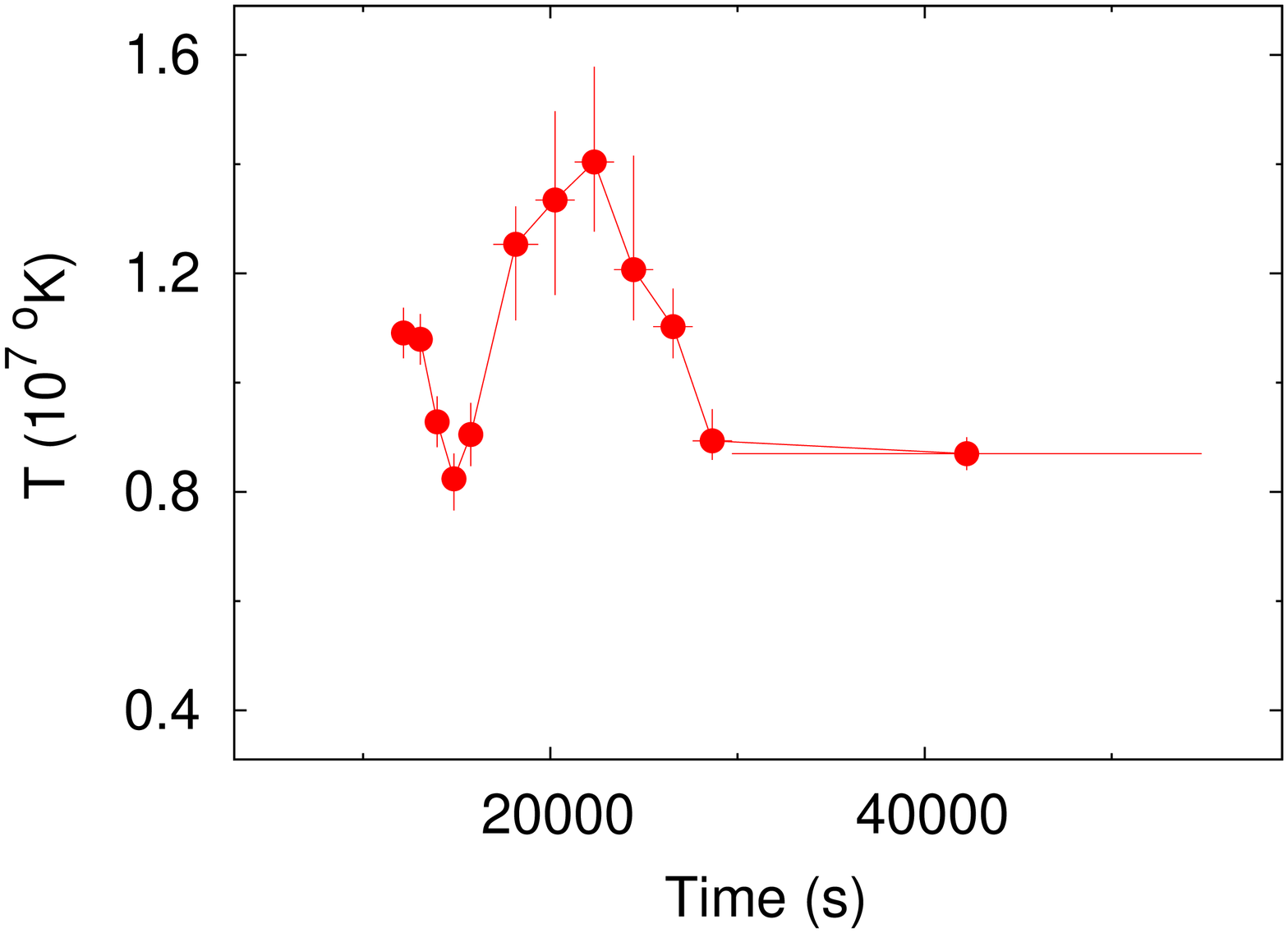}}
\subfigure[]{\includegraphics[width=1.0in,angle=-90]{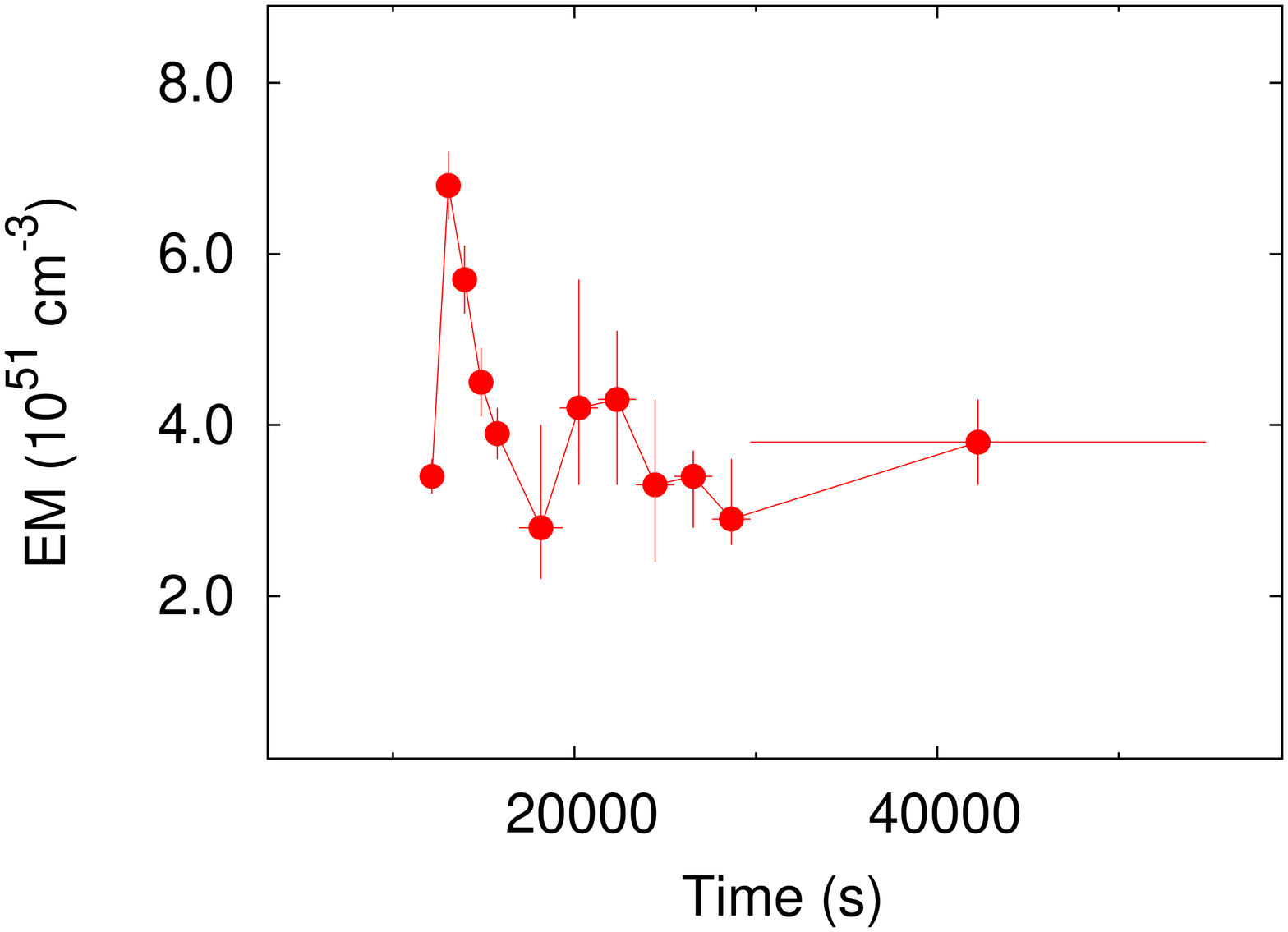}}
\subfigure[]{\includegraphics[width=1.0in,angle=-90]{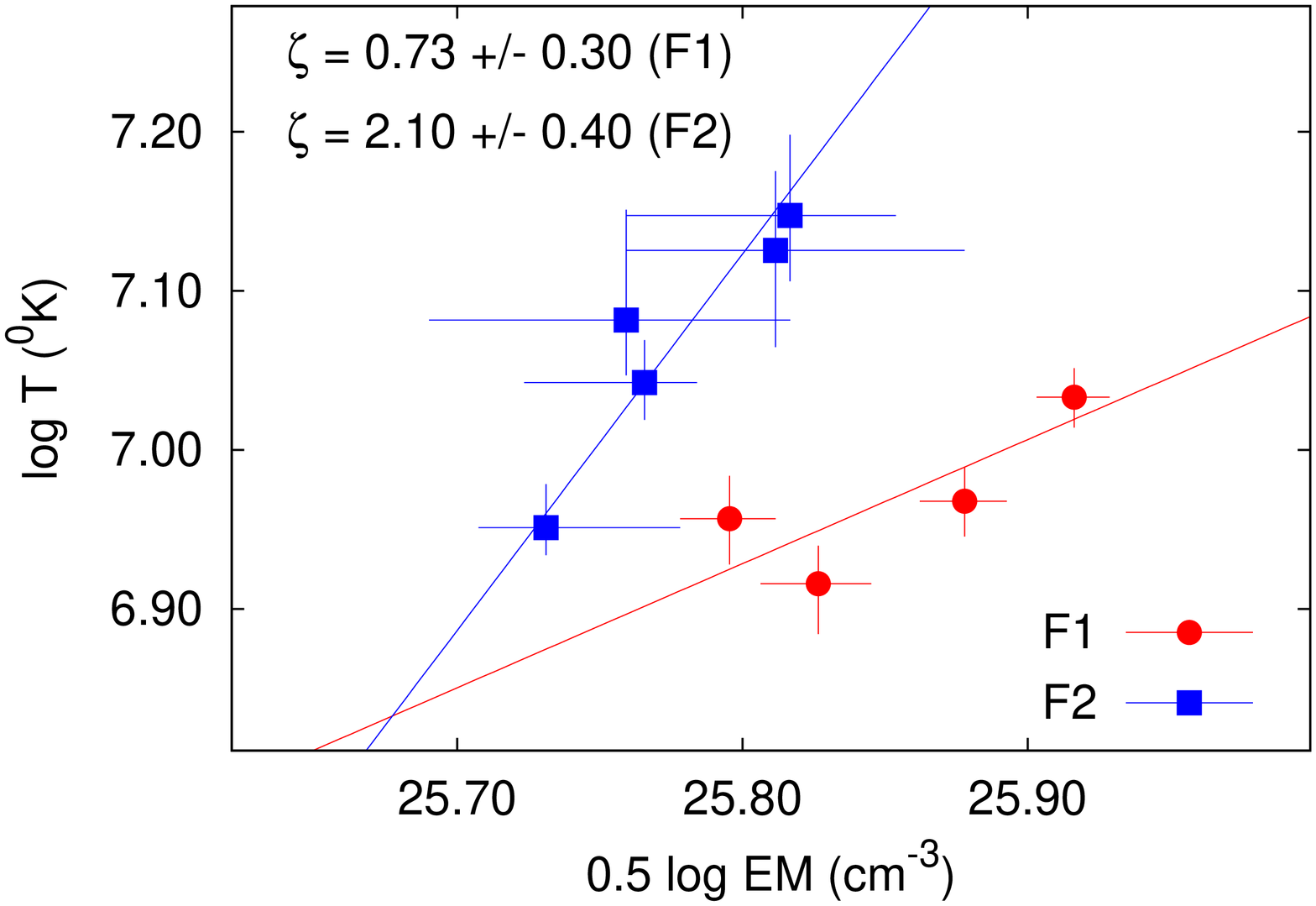}}
\caption{(a) Evolution of temperature, and  (b) emission measure during the flare F1 and F2 and (c) the density–temperature diagram, where EM$^{1/2}$ has been used as a proxy of density.}
\label{fig:fevol}
\end{center}
\end{figure}

   To study the flare emission only, we have performed 1T spectral fits of
the data, with the quiescent emission taken into account by including
its best-fitting 2T model as a frozen background contribution. This is
equivalent to consider the flare emission subtracted of the quiescent
level, allows us to derive one ‘effective’ temperature and one EM
of the flaring plasma. The abundances were kept fix to that of the
quiescent emission. Figs. \ref{fig:fevol} (a) and \ref{fig:fevol} (b) show the temporal evolution of the temperature and corresponding EM of the flares. Both the temperature and EM show the well-defined trends i.e. the changes in the temperature and the EM are correlated with the variations observed in the light curves during the flares. The hardness ratio can reveal the information about temperature variations.  The hardness ratio also varied in the similar fashion to their light curves (see also \cite{pandey08}.

\section{Determinations of density using RGS spectra}
\begin{figure}
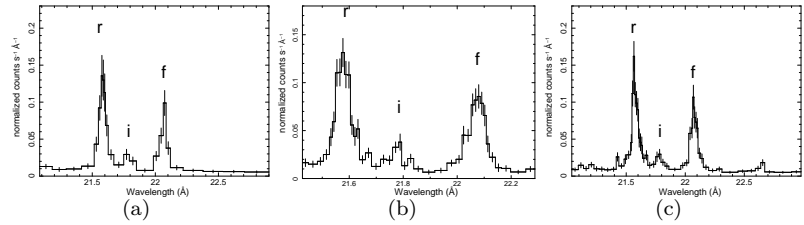

\begin{center}
\subfigure[]{\includegraphics[width=1.0in,angle=-90]{Q_rif.ps}}
\subfigure[]{\includegraphics[width=1.0in,angle=-90]{F1_rif.ps}}
\subfigure[]{\includegraphics[width=1.0in,angle=-90]{F2U_rif.ps}}
\caption{RGS spectra of $\xi$ Boo during quiescent state Q (a), flare state F1 (b), and flare state F2 (c). }
\label{fig:ovii}
\end{center}
\end{figure}

Figs. \ref{fig:ovii} (a), \ref{fig:ovii} (b) and Fig. \ref{fig:ovii} (c) show the    He-like triplets from {\sc O vii} during the quiescent state, flare state F1 and flare state F2+U.
Line fluxes and positions were measured using the XSPEC package by fitting simultaneously the RGS1 and RGS2 spectra with a sum of narrow Gaussian emission lines convolved with the response matrices of the RGS instruments. The continuum emission was described using Bremsstrahlung models at the temperatures of the plasma components inferred from the analysis of EPIC-MOS1 data. The emission measure derived from the analysis of the EPIC data  was used to freeze the continuum normalization. The detailed study  of the RGS spectra was done by Nordon \& Behar \cite{nordon08}.  Therefore, for  the present purpose, we give the line fluxes of He-like triplets from {\sc O vii} in Table 1.

\begin{table}
\tbl{Observed {\sc O vii} line fluxes during the Q, F1 and F2+U states}
{\begin{tabular}{@{}cccccc@{}}
\toprule
\multicolumn{2}{c}{Quiescent (Q)} & \multicolumn{2}{c}{Flare (F1)}&\multicolumn{2}{c}{Flare (F2+U)}\\
$\lambda (\AA)$ & Flux$^a$  & $\lambda (\AA)$ & Flux$^a$ & $\lambda (\AA)$ & Flux$^a$ \\\colrule
22.08(f)&$2.1\pm0.2$&22.07 &$2.3\pm0.2$ & 22.08(f)&$2.2\pm0.3$\\
21.78(i)&$0.8\pm0.2$&21.77 &$0.9\pm0.2$ & 21.77(i)&$0.9\pm0.2$\\
21.58(r)&$3.0\pm0.3$&21.56 &$3.5\pm0.3$ & 21.57(r)&$3.4\pm0.3$\\\botrule
\end{tabular}}
\begin{tabnote}
$^{\text a}$ Measured flux in $10^{-4}$ photons/cm$^2$/s\\
\end{tabnote}
\label{tab:flux}
\end{table}

\begin{figure}
\begin{center}
\subfigure[]{\includegraphics[width=1.9in]{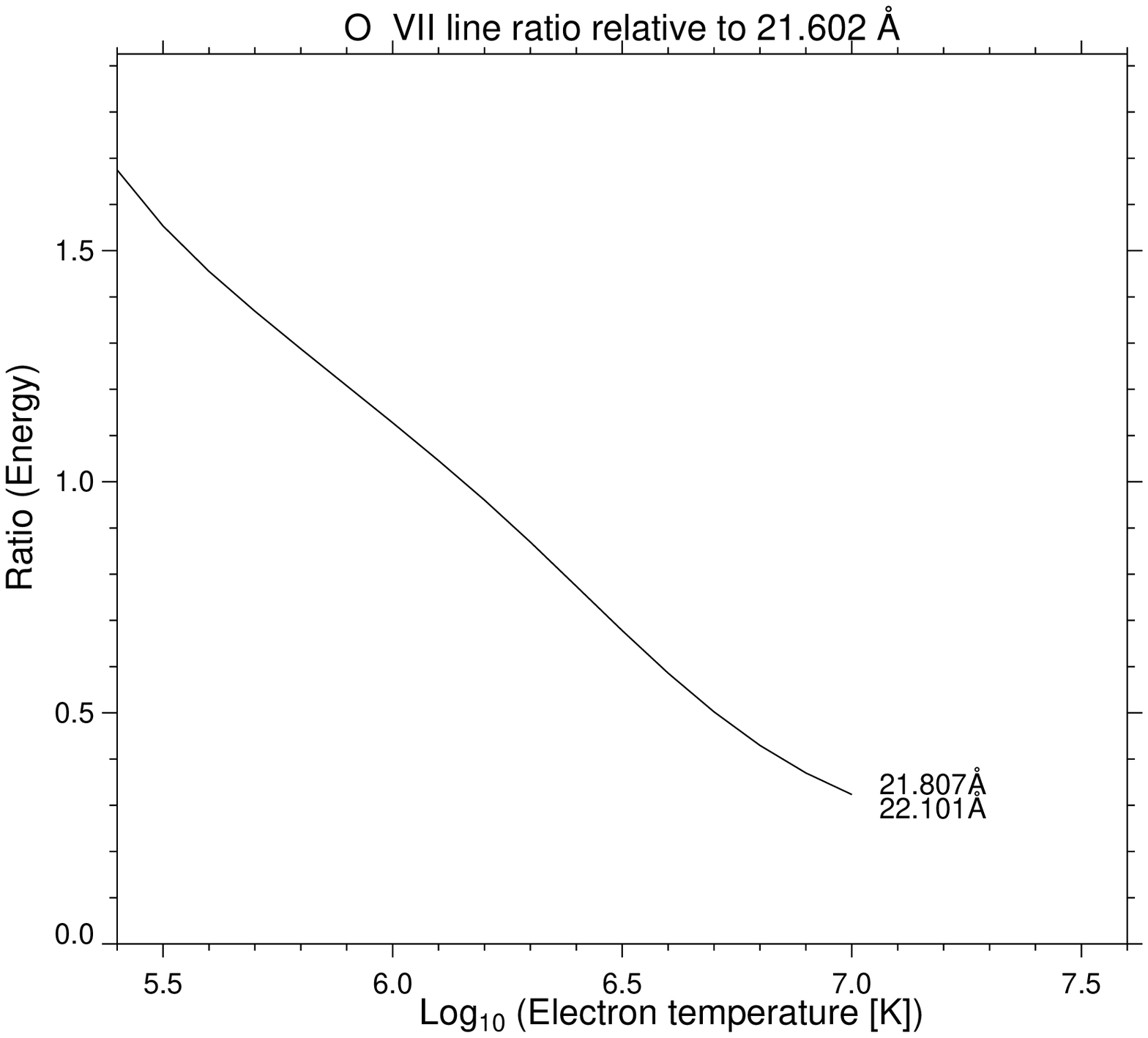}}
\subfigure[]{\includegraphics[width=1.9in]{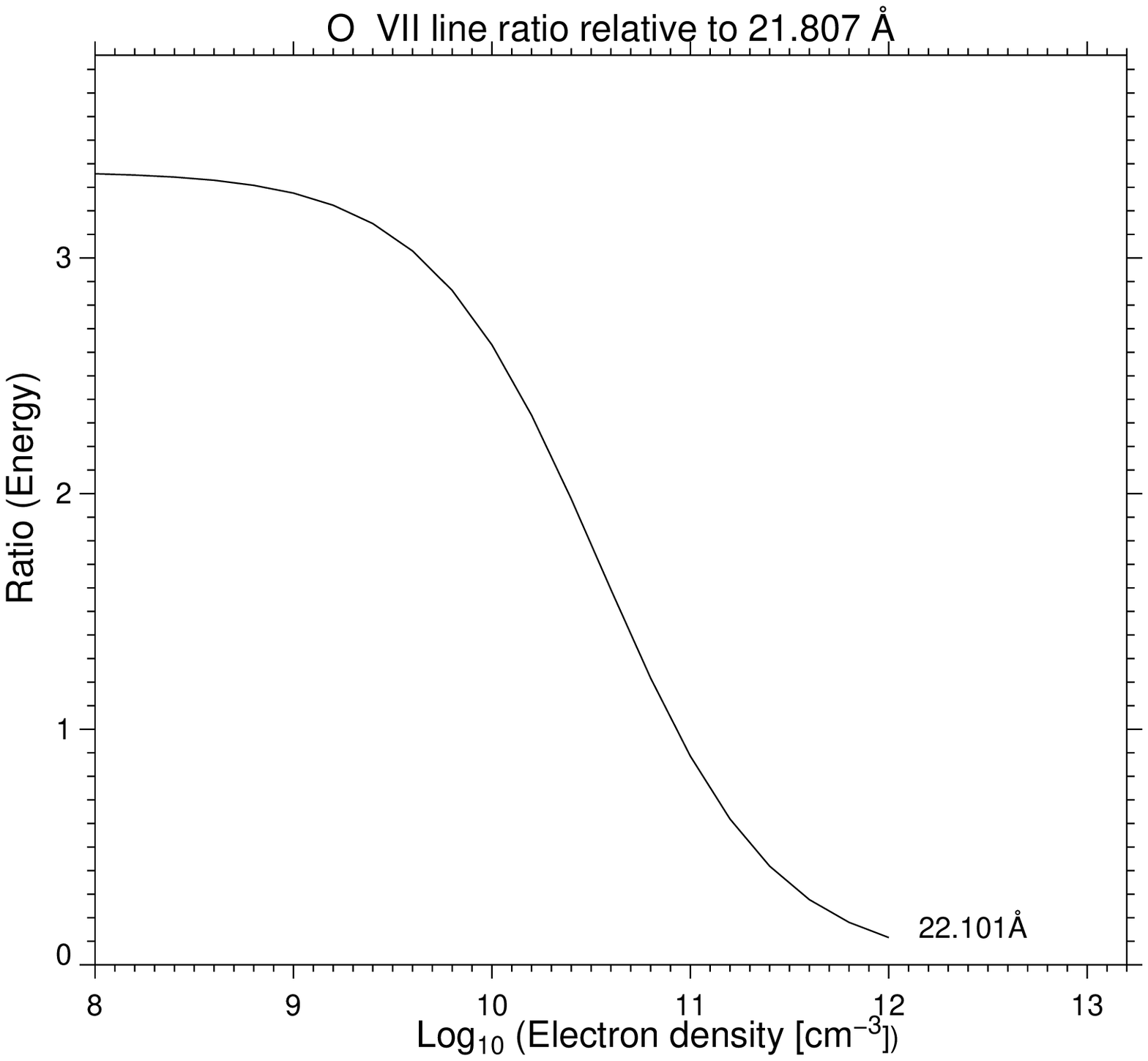}}
\caption{(a) The ratio G = (i + f )/r of the summed intensities of the {\sc OVii} intercombination and forbidden lines over the intensity of the recombination line ($\lambda$= 21.60 \AA) and (b)  intensity ratio R = f /i of the {\sc OVii} forbidden ($\lambda$ = 22.12 \AA) and intercombination ($\lambda$ = 21.80 \AA) lines as a function of electron density,  calculated using the CHIANTI database.}
\label{fig:temp_dens}
\end{center}
\end{figure}

The He-like transitions, consisting of the resonance line (r) 1s$^2$ ~$^1$S$_0$ - 1s2p $^1$p$_1$, the intercombination line (i)  1s$^2$ $^1$S$_0$ - 1s.2p $^3$P1, and forbidden line (f) 1s$^2$ $^1$S$_0$ - 1s.2s $^3$S$_1$ are density- and temperature-dependent. The intensity ratio G = (i+f)/r varies with temperature and the ratio R =f/i varies with electron density due to collisional coupling between the metastable $2^3$S upper level of forbidden line and the $2^3$P upper level of intercombination line. In the RGS wavelength, the {\sc O vii} lines are clean, resolved and potentially suited to diagnose electron density and temperature (see Fig. \ref{fig:ovii}). These lines at 21.60\AA (r), 21.80\AA(i) and 22.10\AA(f) have been used to obtain temperature and density values from the G- and R-ratio using CHIANTI database (version 5.2.1; \cite{landi06}). The  G-ratio as function of temperature and intensity ratio R as a function of density are shown in Figs. \ref{fig:temp_dens}(a) and \ref{fig:temp_dens}(b), respectively. It is clear from Fig. \ref{fig:temp_dens}(a) the temperature was found to be $\sim 2$ MK. Therefore, for  the representative temperature around 2MK, we derive electron densities of $n_e = 1.3_{-0.5}^{+1.2} \times 10^{10}$ cm$^{-3}$ for quiescent state (Q) and $1.0_{-0.8}^{+1.5} \times 10^{10}$ cm$^{-3}$ during flare F1 and $1.6_{-0.8}^{+1.2}\times10^{10}$ cm$^{-3}$ for flare state (F2+U). The electron densities were found well with in a $\sigma$ level for each three states. These densities are found to be similar to that of solar conditions.

\section{Loop modeling}
In a star, flares cannot be resolved spatially. However, by an analogy with solar flares and using flare loop models, it is possible to infer the physical size and morphology of the loop structures involved in a stellar flare. Based on quasi static radiative and conductive cooling during the early phase of the decay. Haisch \cite{haisch83} suggested an approach to model a loop. Given an estimate of two measured quantities, the emission measure (EM) and the decay timescale of the flare ($\tau_d$ ), this approach  leads to the following expression for the loop length ($L_{ha}$):

\begin{equation}
L_{ha}(cm) = 5 \times 10^{-6} (EM)^{1/4} \tau_d^{3/4}
\label{eq:loop_haisch}
\end{equation}

After Haisch \cite{haisch83} various stellar flare models came up. These are (i) two-ribbon flare method \cite{kopp84}, (ii) quasi-static cooling method \cite{van89}, (iii) pure radiation cooling method \cite{pallavicini90}, (iv) rise and decay time method \cite{hawley95}, and  (v) hydrodynamic method \cite{reale97},\cite{reale07}. The two-ribbon flare model assumes that the flare decay is entirely driven by heating released by magnetic reconnection of higher and higher loops and neglects completely the effect of plasma cooling. The other four methods are  based on the cooling of plasma confined in a single flaring loop. The hydrodynamic model includes both plasma cooling and the effect of heating during flare decay. Reale et al. \cite{reale97} presented a method to infer the geometrical size and other relevant physical parameters of the flaring loops, based on the decay time and on evolution of temperature and the EM during the flare decay.  The thermodynamic loop decay time can be expressed \cite{serio91} as

\begin{equation}
\tau_{th} = \frac{3.7\times10^{-4} L}{\sqrt{T_{max}}}
\label{eq:real}
\end{equation}

\noindent
where L is the loop half-length in cm, and T$_{max}$ is the flare maximum temperature (K) calibrated for EPIC instruments as,
\begin{equation}
T_{max} = 0.13 T_{obs}^{1.16}
\label{eq:tmax}
\end{equation}

\noindent
 T$_{obs}$ is the maximum best-fit temperature derived from single temperature fitting of the data. The ratio of the observed exponential light curve decay time $\tau_d$ to the thermodynamic decay time $\tau_{th}$ can be written as a function which depends on the slope $\zeta$ of the decay in the density-temperature plane. For the XMM–Newton EPIC spectral response the ration ${\tau_d}/{\tau_{th}} = F(\zeta)$ is given as \cite{reale07}
\begin{equation}
\frac{\tau_d}{\tau_{th}} = F(\zeta) = \frac{0.51}{\zeta - 0.35} + 1.36 ~~~~~~~~~~~\rm{ (for ~0.35 < \zeta \leq 1.6)}
\label{eq:fzeta}
\end{equation}

\noindent
Combining above equations the expression for semi-loop length is

\begin{equation}
L = L_{hyd} = \frac{\tau_d \sqrt{T_{max}}}{3.7\times10^{-4} F(\zeta)}
\label{eq:loop_decay}
\end{equation}

Alternatively, Reale \cite{reale07} derive the
semiloop length from the rise phase and peak phase of the flare as

\begin{equation}
L_{hyr} = 3\times10^{5/2}t_{M} \frac{T_{max}^{5/2}}{T_M^2}
\label{eq:loop_rise}
\end{equation}

\noindent
$T_M$ and $t_M$ are temperature and time at which density peaks.

Based on  magneto-hydrodynamic simulations Shibata \& Yokoyama \cite{shibata02} assumed that, to maintain stable flare loops, the gas pressure of the evaporated plasma
must be smaller than the magnetic pressure
\begin{equation}
\rm 2nkT \leq \frac{B^2}{8 \pi}
\label{loop_bal}
\end{equation}

\noindent
where B is the minimum magnetic field necessary to confine the flaring plasma in the loop.  The derived equations for B and $L_p$ are

\begin{equation}
\rm  B = 50 \left( \frac{EM}{10^{48}~cm^{-3}} \right) ^{-1/5}  \left( \frac{n_e}{10^{9}~cm^{-3}} \right) ^{3/10}  \left( \frac{T}{10^{7}~K}  \right) ^{17/10} ~G
\label{eq:mag}
\end{equation}

\begin{equation}
\rm L_p = 10^9 \left( \frac{EM}{10^{48}~cm^{-3}} \right) ^{3/5}  \left( \frac{n_e}{10^{9}~cm^{-3}} \right) ^{-2/5} \left( \frac{T}{10^{7}~K}  \right) ^{-8/5} ~cm
\label{eq:loop_pres}
\end{equation}

\begin{table}
\tbl{Loop parameters obtained from various loop models.}
{\begin{tabular}{@{}lccccccc@{}}
\toprule
Flare & $L_{ha}$       & $L_{hyd}$     & $L_{hyr}$    & $L_p$         & B   &  n$_e$ \\
      & ($10^{10}$cm)  & ($10^{10}$cm) &($10^{10}$cm) & ($10^{10}$cm) & (G) & (10$^{10}$cm$^{-3}$) \\\colrule
F1 &$2.2\pm0.2$  & $1.9\pm0.8$ & $1.7\pm0.2$ & $5.7\pm1.4$ &$18.3\pm3.4$& $1.0^{+1.5}_{-0.8}$ \\
F2 &$3.8\pm0.2$ & $7.9\pm0.6$ & $3.6\pm0.8$ & $3.6\pm0.9$ &$29.4\pm6.0$ & $1.6^{+1.2}_{-0.8}$\\\botrule
\hline
\end{tabular}}
\label{tab:loop}
\end{table}

The average temperatures of the loops, usually lower than the real
loop maximum temperatures, are found from the spectral analysis
of the data. According to equation (\ref{eq:tmax}), the
loop maximum temperatures for the flare F1 and  F2  are found to be
$18.9\pm0.8$ and $25\pm2$ MK, respectively. The peak values of temperature are similar to those found in the solar flares. However, the peak emission measure was found to be an order more than those found in the solar flares.
The path of density-temperature (n-T) diagram is shown in Fig. \ref{fig:fevol}(c). The EM$^{1/2}$ was taken as a proxy of density. The solid lines represent the best linear fit to the corresponding data, providing the slope $\zeta$. The values of $\zeta$ for the flare F1 and F2 were found to be $0.7\pm0.3$ and $2.1\pm0.4$, respectively. This indicates that the  flare F1 is driven by the time-scale of the heating process, whereas sustained heating is negligible during the decay of the flare F2.

The semi-loop lengths  derived from above methods are given in table \ref{tab:loop}.
The loop length derived for flare F1 is consistent using three methods (see equations (\ref{eq:loop_haisch}), (\ref{eq:loop_decay}) and (\ref{eq:loop_rise})). However, the loop length derived from pressure balance method ( see equation \ref{eq:loop_pres}) was found to be $\sim$ 3 times  more than other three methods.
The value of $\zeta$ for the flare F2 is outside the domain of the validity of the hydrodynamic method (see equation\ref{eq:loop_decay}), therefore, the loop length derived from hydrodynamic method based on decay phase may not be actual one.  The loop length derived from other three methods (see equations (\ref{eq:loop_haisch}), (\ref{eq:loop_rise}) and (\ref{eq:loop_pres})) is found to be consistent at a value of  $3.6\times10^{10}$ cm. The magnetic field derived using the equation (\ref{eq:mag}) is 18.9 and 29.4 Gauss for the flares F1 and F2, respectively.

The total energy rate is defined as $H = E_H\times V$, where $E_H = 10^{-6}T_{max}^{3.5}L^{-2}$ erg s$^{-1}$ cm$^{-3}$ \cite{rosner78} heating rate per unit volume and $V = EM/n^2$ volume of the loop. Therefore, the total energy rate was determined to be $5.4\times10^{30}$ and $9.8\times10^{30}$ ergs s$^{-1}$ for the flare F1 and F2, respectively. The peak X-ray luminosity  during the flare F1 and F2 was found to be $1.02\times10^{29}$  and $1.11\times10^{29}$ ergs s$^{-1}$, which is only 2 and 11 \% of the total energy rate, respectively.

\section{Search for coronal oscillations}
\begin{figure}
\begin{center}
\includegraphics[width=1.5in,angle=90]{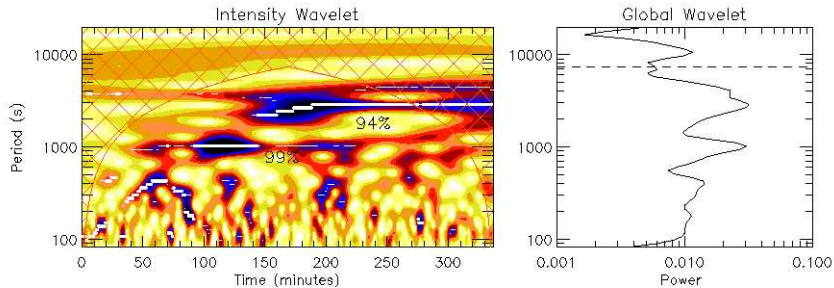}
\caption{The wavelet power spectrum of the U part of X-ray light curve of $\xi$ Boo.}
\label{fig:wavelet}
\end{center}
\end{figure}

   We have used the wavelet analysis IDL code "Randomlet"
developed by E. O'Shea. The program executes a randomization
test \cite{linnell85}, \cite{oshea01} which
is an additional feature along with the standard wavelet analysis
code \cite{torrence98} to examine the existence of
statistically significant real periodicities in the time series data.
The advantage of using this test is that it is distribution free or
nonparametric, i.e., it is not constrained by any specific noise
model such as Poisson, Gaussian, etc. Using this technique,
many important results have been published by analyzing
approximately evenly sampled data (e.g., \cite{oshea01}, \cite{srivastava08a}, \cite{srivastava08b}).
   The wavelet power transforms of the U part of the X-ray light
curve of cadence 40 s are shown in the  left panel of Fig. \ref{fig:wavelet}, where the darkest regions show the most enhanced oscillatory power in
the intensity wavelet spectrum. The crosshatched areas are the
cone of influence (COI), the region of the power spectrum where
edge effects, due to the finite lengths of the time series, are likely
to dominate. The maximum allowed period from COI, where
the edge effect is more effective, is 7363 s. In our wavelet
analysis, we only consider the power peaks and corresponding
real periods (i.e., probability $> 95$\%) below this threshold.
The period with maximum power detectable outside the COI
is 2883 s with a probability of 94\% and the repetition of
only 3 cycles in the time series. However, another peak is
visible at a period of 1019 s with a probability of 99\% and
the repetition of 11 cycles. Therefore, only the periodicity of
1019 s satisfies the recently reported stricter criteria of O'Shea
\& Doyle\cite{oshea09}, i.e., at least four cycles of repetition over the
lifetime of the oscillations. The right panel in Fig.\ref{fig:wavelet}
shows the global wavelet power spectra of the time series from
which the statistically significant period (1019 s) is selected.
Wavelet analysis was also performed in the decay phase of the flare F2, and no significant periodicity was found.

\section{Evidence of Fast-kink Oscillation}
 In this section, we probe the presence of various MHD modes associated with  the flare F2  observed in $\xi$ Boo.  The overdense magnetic loops are pressure balanced structures and may contain fast kink and sausage oscillation modes with phase speed greater than the local Alfv\'enic speed ($\omega/k > v_{A}$), and slow acoustic modes with sub-Alfv\'enic speed ($c_{0} < v_{A}$) under coronal conditions.  The expressions for the oscillation periods of slow ($P_{s}$), fast-kink ($P_{fk}$), and fast-sausage ($P_{fs}$) fundamental modes are given as follows (\cite{edwin83}, \cite{roberts84}, \cite{aschwanden99}) :

\begin{equation}
P_{s} = \frac{2 L}{jc_{T}}=\frac{2 L}{j c_{o}}\left[1 + \left(\frac{c_{o}}{v_{A}}\right)^{2}\right]^{1/2}
\approx 1300 \frac{L_{10}} {\sqrt T_{6}} s
\label{eq:s}
\end{equation}

\begin{equation}
P_{fk} = \frac{2L}{jc_{k}} = 4\pi^{1/2}\frac{L}{j}\left( \frac{\rho_{o} + \rho_{e}}{B_{o}^{2} + B_{e}^{2}}\right)^{1/2}
\approx 205 \frac{L_{10} \sqrt{n_{9}}}{B_{10}} s
\label{eq:fk}
\end{equation}

\begin{equation}
P_{fs} = \frac{2\pi a}{c_{k}} = 4\pi^{3/2}a\left( \frac{\rho_{o} + \rho_{e}}{B_{o}^{2} + B_{e}^{2}}\right)^{1/2}
\approx 6.4 \frac{a_{8} \sqrt{n_{9}}}{B_{10}} s
\label{eq:fs}
\end{equation}

~~~\\

\noindent
where $L$, $a$, $\rho$ and $B$ are loop length, loop width, ion mass density and magnetic field,
respectively. The $c_{T} = {c_{o} v_{A}}/ {(c_{o}^2+v_{A}^2)}^{1/2}$, where $c_{o}$ and $v_{A}$
are sound  and Alfv\'enic speeds respectively. The $c_{k}$ is phase speed. Under coronal conditions, $B_{o} \approx B_{e}$,
$\rho_{e} << \rho_{o}$, $ n_H \approx n_e$, and $c_{o} << v_{A}$. The subscripts $'o'$ and $'e'$
stands for 'inside', and 'outside' of the loop. The $j$ is the number of nodes in fast-kink and slow oscillations,
and is equal to 1 for the fundamental mode. The approximate relation in equations (\ref{eq:s}) - (\ref{eq:fs}) are given in coronal conditions
for the fundamental oscillation periods of different modes in magnetic loops. Therefore, these approximate
equations are also valid for the stellar loops in the coronae of magnetically active Sun-like stars.
The parameters are expressed in units of $L_{10} = L/10^{10} cm $, $a_{8} = a/10^{8} cm$, $n_{9} = n_o/10^{9} cm^{-3}$, $T = T_{e}/10^{6} K$, and $B_{10} = B/10 G$.

Using observationally estimated loop length ($3.6\pm0.8\times 10^{10}$ cm) and width($a = 0.1L$), density ($1.6_{-0.8}^{+1.2} \times 10^{10}$ cm$^{-3}$) and magnetic field ($29.4\pm6.0$G) for flare F2,
and equations (\ref{eq:s})-(\ref{eq:fs}), the estimated oscillations  periods $P_s$, $P_{fk}$ and $P_{fs}$ for fundamental modes are found to be $1586\pm353$,  $1004\pm391$ and $313\pm121$ s, respectively.
The observationally derived oscillation period $\sim1019$  s approximately
matches with the theoretically derived oscillation period  for fundamental fast kink mode. These oscillations may be formed either by the superposition of oppositely propagating
fast-kink waves or due to flare generated disturbances near the
loop apex. The flaring activity may be the possible mechanism of
the generation of such oscillations in the stellar loops.
 Cooper et al.\cite{cooper03} have found that the transverse kink modes
of fast magnetoacoustic waves may modulate the emission
observable with imaging telescopes despite their incompressible
nature. This is possible only when the axis of an oscillating loop
is not orientated perpendicular to the observer's line of sight. Using this theory O'Shea et al.\cite{oshea07} have also found the first two harmonics in kink oscillations in cool solar loops. Similar effect may also cause the modulation in the emission from stellar loops.
   Resonant absorption is the most efficient mechanism theorized for the damping of a kink mode in which energy of the mode is transferred to the localized Alf\'evanic oscillations of the inhomogeneous layers at the loop boundary \cite{lee86}. Intensity oscillations due to the fast-kink mode of magnetoacoustic waves do not show any decay of the amplitude during the initial 250 minutes in the post-flare phase observations of $\xi$ Boo. This is probably due to an impulsive driver, which forces such kind of oscillations and dominates over the
dissipative processes during the flare F2 of $\xi$ Boo.

\section{Summary}
 The observed flares in G-dwarf $\xi$ Boo are similar to the solar arcade flares, which are as strong as those in M-dwarfs and are much smaller than the flare observed in dMe star, giants and pre-main-sequence analogous. The loop lengths are found to be  smaller than the stellar radius. The electron densities during the flares and quiescent state were found to be of the order of 10$^{10}$ cm$^{-3}$. We also report the first observational evidence of fundamental fast-kink oscillations in stellar loops during a post-flare phase of heightened emission in the cool active star $\xi$ Boo. Based on the solar analogy, such observations may shed new light on the dynamics of the magnetically structured stellar coronae and its local plasma conditions.

\end{document}